\documentclass[aps,prl, twocolumn,amsmath,amssymb,superscriptaddress]{revtex4-2}% APS journal style
\usepackage{mathtools}
\usepackage{amsmath,amssymb,amsfonts,bm}
\usepackage{graphicx}
\usepackage{xcolor}
\usepackage{bbold}
\usepackage{graphicx}
\usepackage{dcolumn}% 
\usepackage{epstopdf}
\usepackage{epsfig}
\usepackage[caption=false]{subfig}
\usepackage{url}
\usepackage{hyperref}
\usepackage{verbatim}
\usepackage[export]{adjustbox}
\usepackage{scalerel}
\usepackage{braket}
\usepackage{csquotes}
\usepackage{mathrsfs}
\usepackage[normalem]{ulem}
\usepackage{amsthm}
\usepackage{enumitem}
\usepackage{multirow}
\usepackage{booktabs}
\usepackage{pifont}

\usepackage{colortbl}

\newcommand{\average}[1]{\left\langle #1 \right\rangle}
\def\avec{\mathbf{a}}
\def\pvec{\mathbf{p}}

\graphicspath{{FirstRoundPRL_v3_14.04.25}}

\usepackage{ifthen}

\newcommand{\be}{\begin{equation}}
\newcommand{\ee}{\end{equation}}
\newcommand{\ba}{\begin{aligned}}
\newcommand{\ea}{\end{aligned}}
\newcommand{\bea}{\begin{eqnarray}}
\newcommand{\eea}{\end{eqnarray}}
\newcommand{\bmult}{\begin{multline}}
\newcommand{\emult}{\end{multline}}

\newcommand{\titleinfo}{Measurement-Induced Phase Transition in State Estimation of Chaotic Systems and the Directed Polymer}

\begin{document}

\title{\titleinfo}

\author{Federico Gerbino}
\affiliation{Laboratoire de Physique Théorique et Modèles Statistiques, Université Paris-Saclay, CNRS, 91405 Orsay, France}

\author{Guido Giachetti}
\affiliation{Laboratoire de Physique de l'\'Ecole Normale Sup\'erieure, CNRS, ENS $\&$ PSL University, Sorbonne Universit\'e, Universit\'e Paris Cité, 75005 Paris, France}

\author{Pierre Le Doussal}
\affiliation{Laboratoire de Physique de l'\'Ecole Normale Sup\'erieure, CNRS, ENS $\&$ PSL University, Sorbonne Universit\'e, Universit\'e Paris Cité, 75005 Paris, France}

\author{Andrea De Luca}
\affiliation{Laboratoire de Physique Th\'eorique et Mod\'elisation, CY Cergy Paris Universit\'e, \\
\hphantom{$^\dag$}~CNRS, 95302 Cergy-Pontoise, France}

\begin{abstract}

We introduce a solvable model of a measurement-induced phase transition (MIPT) in a deterministic but chaotic dynamical system with a positive Lyapunov exponent. In this setup, an observer only has a probabilistic description of the system but mitigates chaos-induced uncertainty through repeated measurements. 
Using a minimal representation via a branching tree, we map this problem to the directed polymer (DP) model on the Cayley tree, although in a regime dominated by rare events. By studying the Shannon entropy of the probability distribution estimated by the observer, we demonstrate a phase transition distinguishing a chaotic phase with reduced Lyapunov exponent from a strong-measurement phase where uncertainty remains bounded. 
Remarkably, the location of the MIPT transition coincides with the freezing transition of the DP, although the critical properties differ. We provide an exact, universal scaling function describing the entropy growth in the critical regime. Numerical simulations confirm our theoretical predictions, highlighting a simple yet powerful framework to explore measurement-induced transitions in classical chaotic systems.
\end{abstract}

\maketitle
\date{\today}

\textit{Introduction ---}
In recent years, much attention has been devoted to the interplay between unitary evolution and external noise~\cite{preskill2018quantumcomputingin,Hoke2023, PhysRevX.8.021014, PhysRevX.13.011043,PhysRevX.13.011045, Bernard_2021, roy2, lunt2022quantumsimulationusing,botzung2021engineereddissipationinduced}, particularly the one induced by the action of quantum measurements~\cite{PhysRevA.36.5543}. While a closed quantum system tends to encode local information into nonlocal degrees of freedom, resulting in the production of entanglement entropy~\cite{dalessio2016,RevModPhys.91.021001, PhysRevX.7.031016, PhysRevX.8.041019, PhysRevX.9.021033}, measurements of local quantities interfere by extracting information from the quantum state. As a result of such competition, two phases can emerge~\cite{PhysRevX.9.031009, PhysRevB.98.205136, PhysRevB.101.104302}: at weak measurements, non-local quantum correlations do grow (volume law)~\cite{PhysRevX.10.041020, PhysRevB.100.134306}, and the system thus exhibits a spontaneous error-correcting capacity~\cite{PhysRevLett.125.030505, fan2021selforganizederror,li2021statisticalmechanicsof};  conversely, at strong measurements, information remains confined to local degrees of freedom and entanglement growth saturates (area law), allowing for an effective matrix-product-states description~\cite{PhysRevLett.91.147902, PhysRevResearch.6.033220}. Experiments have confirmed this phenomenology~\cite{Koh2023,czischek2021simulating,noel2022measurementinducedquantum}, although the observability of this transition in extended systems has been questioned because of the burden of post-selection~\cite{PhysRevLett.125.070606,ippoliti2021postselectionfreeentanglement}.

Due to the inherently random outcome of quantum measures, monitored systems can be seen as disordered with time being an extra spatial dimension. However, the distribution of outcomes is not fixed a-priori but is determined by the state itself (according to Born's rule). This implies that the results of measurements at different times are strongly correlated (up to the extreme case of the Zeno effect where they do not change)~\cite{PhysRevB.100.134203, PhysRevLett.125.070606}, in contrast with usual disordered systems where short-range disorder correlations are generally assumed~\cite{mezard1987spin,A_J_Bray_1980}. %Not surprisingly, an accurate characterization of the critical point is difficult. 
Nonetheless, the interpretation as disordered systems for monitored non-interacting fermions~\cite{10.21468/SciPostPhys.7.2.024, PhysRevB.105.094303, PhysRevLett.126.170602, PhysRevX.11.041004, PhysRevA.107.032215, Fidkowski2021howdynamicalquantum, PhysRevB.108.L020306, merritt2023entanglementtransitionswith}, has revealed similarities with the Anderson transition~\cite{evers2008andersontransitions} and non-linear sigma models~\cite{PhysRevX.13.041045, PhysRevX.13.041046}.
In general, approaches based on annealed averages~\cite{PhysRevLett.127.140601, PhysRevB.105.L241114, PhysRevB.103.224210,bao2020theoryofthe, zabalo2020criticalpropertiesof,zhang2021emergentreplica}, mean field~\cite{PRXQuantum.2.010352, bentsen2021measurement, PhysRevB.102.064202,giachetti2023elusivephasetransitionreplica}, random matrices~\cite{deluca2024universalityclassespurificationnonunitary, quantum6020016}, and extensive numerics~\cite{tang2020measurementinducedphase,PhysRevResearch.6.033220, PhysRevB.109.174307, turkeshi2022measurementinducedcriticality} provided the general picture that interacting quantum systems in $d+1$ dimensions undergo a MIPT, with a critical point described by a yet unknown nonunitary conformal field theory~\cite{PhysRevB.109.014303, PhysRevLett.128.050602,PhysRevB.108.104203,li2021conformal}. 

\begin{figure}
    \centering
    \includegraphics[width=\columnwidth]{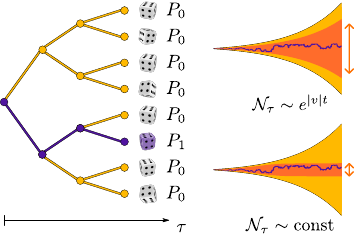}
    \caption{ \textit{Left:} Cayley tree ($K=2$), modeling the exponential growth $\Delta x_{\rm max} \sim e^{\lambda t}$ of the uncertainty on the position of a particle. 
    The purple branch depicts the true trajectory $x_\tau$ of the particle: at each time-step $\tau$, a finite-precision measurement on each site is performed, whose outcomes $a_j$ are distributed with $P_{n_j}(a_j)$ depending on the occupancy of the site $j$, $n_j = 0,1$. \textit{Right:} sketch of the two phases of the model. The orange region represents a subregion of size $\mathcal{N}_\tau \sim e^{S_t}$ in which the observer can be reasonably sure the particle is located. The effectiveness of measurements determines whether $\mathcal{N}_\tau \sim {\rm const}$ or $\mathcal{N}_\tau$ still scale exponentially, but with a reduced Lyapunov exponent $|v| < \lambda$.
    }
    \label{fig:binary_tree}
\end{figure}

A similar protocol can also be considered in a purely classical context, where the stochastic description of a system is continuously updated by measurements according to Bayes' theorem~\cite{PhysRevB.106.024305, pizzi2022bridgingthegap}; in contrast with quantum MIPTs, there is no post-selection barrier since the measurements do not affect the actual state of the system. For a $1D$ diffusive particle undergoing Bayesian monitoring, a short-time KPZ behavior has been suggested~\cite{PhysRevLett.129.260603, kim2024planteddirectedpolymerinferring}. 
In this Letter, we present a simple, solvable model exhibiting a monitoring-induced phase transition. Let us consider a generic chaotic classical dynamical system with positive Lyapunov exponents. Although it is a deterministic system, at the practical level, the exponential growth of any finite uncertainty about the initial conditions allows only a probabilistic description. To mitigate this uncertainty, an observer makes measurements, in order to update their estimate of the state of the system. A minimal description of this setup is given by a branching tree, in which the branching ratio $K$ is related to the (maximum) Lyapunov exponent, see Fig.~\ref{fig:binary_tree}.  

This formulation allows a mapping to the famous DP problem in this geometry~\cite{derrida1988polymers}. In~\cite{kim2024planteddirectedpolymerinferring}, within this setup, the existence of two distinct phases has been demonstrated when assuming knowledge of the exact trajectory. These two phases correspond to a time-integrated overlap between the estimated and exact trajectory being $0$ or $1$. Here, however, consistent with the idea that the true trajectory is not knowable, we study the problem exclusively from the observer's point of view. This has a direct relation  with the quantum case with Bayes theorem replacing Born's rule: the aforementioned correlation between measurements outcomes requires to analyze the DP problem in a regime of rare events~\cite{carpentier2001glass}. We study the Shannon entropy of the observer-estimated probability distribution: we demonstrate the existence of a phase transition, between (i) a regime in which chaos persists albeit with a reduction in the effective Lyapunov exponent, and (ii) a regime in which the uncertainty saturates with time. Interestingly, the location of the transition coincides with that of the freezing transition of the DP. In addition, the Shannon entropy displays universal critical properties in proximity of the transition that we are able to thoroughly characterize by providing an explicit, exact and universal scaling function describing the entropy growth at long times.

\textit{The model ---} 
We consider an abstract setting with an external observer looking at a dynamical system: in order to simulate the chaotic growth of uncertainty, we assume the state of the system to be one of the nodes of a Cayley tree branching with ratio $K$ along the effective discrete-time direction $t = \tau \Delta t$, $\tau \in \mathbb N$. While the evolution is deterministic, i.e., the system follows a specific path connecting the root to a leaf, because of their finite knowledge about the initial condition, the observer can only rely on a probabilistic description. We thus have a particle undergoing ``directed'' random walk: a particle in $x_\tau \in \{j\}_{j=1}^{K^\tau}$ at time $\tau$, hops to one of the $K$ possible states $K (x_\tau-1) + m$  at level $\tau+1$,
with $m\in \{ 1,\dots,K\}$.  Let us notice that, by setting $K = e^{\lambda \Delta t}$, the maximum separation between two trajectories at time $t$ is $\Delta x_{\rm max} \sim e^{\lambda t}$ so that $\lambda$ can be identified with the maximal Lyapunov exponent.  

We denote as $p_j^{(\tau)}$ the probability for the observer that the system is located in $j$ at time-step $\tau$. Right after the random-walk step, the observer performs instantaneous measurements of each site in order to mitigate the growth of uncertainty. In order to measure site $j$ at time-step $\tau$, a measuring apparatus returns a random variable $a_j^{(\tau)}$ distributed with probability density $P_1(a)$ if the site is occupied at time-step $\tau$, and $P_0(a)$ otherwise. 
For clarity, although the discussion is generic, throughout this Letter, we will keep the Gaussian case as a yardstick where $P_0$ and $P_1$ are Gaussian distributions with unit variance and mean $0$ and $\mu$ respectively.
While the outcome is not deterministic, the measurement is classical, so that the system is unaffected by it.
As measurements from different sites are statistically independent, the probability of the outcomes $\mathbf{a^{(\tau)}}\equiv\{a_j^{(\tau)}\}_{j=1}^{K^\tau}$ conditioned to the particle being in site $j$ are
\begin{equation} \label{eq:a_cond_x}
    P(\mathbf a^{(\tau)} |x_\tau = j) =  \frac{P_1(a^{(\tau)}_{j})}{P_0(a^{(\tau)}_{j})}  \ \prod_{j'=1}^{K^\tau} P_0(a^{(\tau)}_{j'}) \, .
\end{equation}
From the knowledge of the measurement outcomes $\mathbf a^{(\tau+1)}$
at level $\tau+1$, the observer can update the estimate of the probabilities from $p_j^{(\tau)}$ to $p_j^{(\tau+1)}$ using Bayes' theorem
\begin{equation} \label{eq:p_evolve}
   p_j^{(\tau+1)} = \frac{P(\mathbf a^{(\tau+1)}|j) \, p_j^{(\tau+1,-)} }{\sum_{j'} P(\mathbf a^{(\tau+1)}|j') \,p_{j'}^{(\tau+1,-)} }  \,,
\end{equation}
where $p_{j}^{(\tau+1, -)} = p_{\lceil j/K \rceil}^{(\tau)}/K $ denotes the probabilities right before the measurements and $\lceil \ldots \rceil$ is the ceiling function.
We will assume that the process is repeated several times, with the true trajectory chosen uniformly among all paths on the tree. For a given true trajectory, Eq.~\eqref{eq:a_cond_x} determines the distribution of outcomes; then, knowing the outcomes, Eq.~\eqref{eq:p_evolve} gives the distribution inferred by the observer. To determine how effective the observer's measurements are, we will be interested in analyzing typical properties of the inferred distribution. 

\textit{Mapping to a directed polymer ---} 
To simplify the dynamics induced by Eqs.~(\ref{eq:a_cond_x}, \ref{eq:p_evolve}) we make two observations. First, since the denominator in Eq.~\eqref{eq:p_evolve} serves only to impose normalization, for a known set of measurement outcomes $\{\mathbf{a^{(\tau')}}\}_{\tau'\leq \tau}$, we introduce a set of non-normalized variables $\mathbf{z}^{(\tau)}:=\{z_j^{(\tau)}\}_{j=1}^{K^\tau}$ evolving with $z^{(\tau+1)}_j=B(a_j^{(\tau+1)}) z_{\lceil j/K\rceil}^{(\tau)}$, where $B(a) := P_1(a)/(K P_0(a))$. As at each step $p_j^{(\tau)} \propto z_j^{(\tau)}$, we can recover $p_j^{(\tau)} = z_j^{(\tau)}/Z^{(\tau)}$, with $Z^{(\tau)} := \sum_{j=1}^{K^\tau} z_j^{(\tau)} = \sum_{\rm paths} \prod_{x \in \rm path} B(a_{x})$. Interpreting $B(a)$ as a Boltzmann weight associated to the nodes of the tree, $Z^{(\tau)}$ can be seen as the partition function of a DP on the tree~\footnote{In the language of the DP, the chemical potential and inverse temperature
are proportional to the mean and to the r.m.s of $\ln P_1(a)/(K P_0(a))$ respectively.}.

Let us denote as $\mathbf{p}^{(\tau)}:= \{p_j^{(\tau)}\}_{j=1}^{K^\tau}$ the full set of probabilities for a given true trajectory $x_\tau$ and outcomes $\{\mathbf{a^{(\tau')}}\}_{\tau'\leq \tau}$.
Consider an arbitrary functional $F[\mathbf{p}^{(\tau)}]$. Then, its average over the uniform distribution of $x_\tau$ and the corresponding distribution of outcomes  $\{\mathbf{a^{(\tau')}}\}_{\tau'\leq \tau}$ reads
\begin{equation}
\label{eq:meanF}
  \average{F[\mathbf{p}^{(\tau)}]} := 
  \frac{1}{K^\tau}\sum_{x_\tau = 1}^{K^\tau}
\int d\mathbf{a} \;
F[\mathbf{p}^{(\tau)}]
\prod_{\tau'=1}^\tau P(\mathbf{a}^{(\tau')} | x_{\tau'})
\end{equation}
where $d\mathbf{a} = \prod_{j,\tau'} da_j^{(\tau)}$ and we implicitly used that on the tree, the end point $x_\tau$ also determines all the intermediate points $x_{\tau'}$ for $\tau' = 1,\ldots, \tau$.
For clarity, the $\mathbf{p}^{(\tau)}$'s in the r.h.s. of Eq.~\eqref{eq:meanF} depend implicitly on $\{\mathbf{a^{(\tau')}}\}_{\tau'}$ through Eq.~\eqref{eq:p_evolve}.
Using Eq.~\eqref{eq:a_cond_x}, we can identify in Eq.~\eqref{eq:meanF}, a factor $\prod_{\tau',j} P_0(a_j^{(\tau')})$, which we interpret as the measure for the variables $a_j^{(\tau')}$. From the definition of $B(a)$, the remaining factor can be recognised as the normalization $Z^{(\tau)}$, leading to~(see also \cite{supplnote})
\begin{equation} \label{eq:bornrule}
    \average{F[\mathbf{p}^{(\tau)}]}%_{\mathbf{a},x_\tau}
    = \average{F\left[\left\{\frac{z^{(\tau)}_j}{Z^{(\tau)}}\right\}_{j}\right]\,Z^{(\tau)}}_0 \,,
\end{equation}
where $\average{ \dots }_0$ denotes the process where all $a_{j}^{(\tau')}$ are i.i.d. with distribution $P_0(a)$.
Note that, by definition, $\langle B(a) \rangle_0 = 1/K$ which consistently implies $\average{Z^{(\tau)}}_0=1$. 

\textit{Entropy dynamics ---} 
The more the two $P_{0,1}(a)$ distributions differ, the more the observer will be able to discern where the particle is, typically resulting in a set of $p_j^{(\tau)}$ peaked around few $j$'s. A possible quantifier of the ``distance" between $P_{0,1}(a)$, and therefore of the effectiveness of the measurement protocol, is the so-called Kullback-Leibler divergence of $P_1(a)$ with respect to $P_0(a)$, $D_{\rm KL} (P_1 \parallel P_0) 
:= \average{\ln(P_1(a)/P_0(a)) }_1 \geq 0$, where 
$\average{\dots }_1= \int da \dots P_1(a)$,
which measures the surprise of an observer to find that $a$ is distributed according to $P_1(a)$, while $P_0(a)$ is expected. If $P_{0,1}$ are taken to be Gaussian with a relative shift $\mu$ one has $D_{\rm KL} (P_1 \parallel P_0) = \mu^2/2$.

Our goal is now to determine whether the observer is able to effectively locate the particle. To this aim, we evaluate the growth in time of the average Shannon entropy $\average{S_t} = \average{-\sum_{j=1}^{K^\tau} p_j^{(\tau)} \ln p_j^{(\tau)}} $
which quantifies the degree of uncertainty on the particle location. More specifically, one can think of $\mathcal{N}_\tau \sim e^{S_t}$ as an estimate of the number of sites in which the probability of finding the particle is significantly different from zero, see Fig.~\ref{fig:binary_tree}.
According to Eq.~\eqref{eq:bornrule} it can be expressed as
\begin{equation} \label{eq:Z_entropy}
    \average{S_t} = - \sum_{j=1}^{K^\tau} \average{z_j^{(\tau)}\ln z_j^{(\tau)}}_0 +  \average{Z^{(\tau)} \ln Z^{(\tau)}}_0  \,.
\end{equation}
The first term in this expression can be promptly computed as each $z_j^{(\tau)}$ is expressed as the product of independent factors leading to $\sum_{j=1}^{K^\tau} \average{z_j^{(\tau)}\ln z_j^{(\tau)}}_0 = v t$ (see End Matter (EM)). Here, we introduced
\begin{equation} \label{eq:v_ref}
    v \equiv \frac{1}{\Delta t} D_{\rm KL} (P_1 \parallel P_0) - \lambda \, , 
\end{equation}
which can be seen as a control parameter accounting for the competition between the measurement precision, quantified by $D_{\rm{KL}}$, and the chaotic spreading of the trajectories characterized by the Lyapunov exponent $\lambda$.
% = \ln K / \Delta t$. 

On the other hand, the second term in the r.h.s. of Eq.~\eqref{eq:Z_entropy} is more involved.  
A standard method in disordered systems to cope with this kind of averaging is to calculate integer moments $\langle Z^n \rangle$ for generic $n$ and subsequently consider an analytic continuation in $n$, the so-called replica trick. The typical value of the free energy $\propto \ln Z$ in quenched-disorder problems amounts to consider the $n \to 0$ expansion. In contrast, in monitoring problems, as we anticipated, the fact that the inference is based on Bayes' theorem~\eqref{eq:p_evolve} and \eqref{eq:bornrule} means that the entropy~\eqref{eq:Z_entropy} is controlled by $Z \ln Z = \partial_n Z^n|_{n=1}$. It follows that the entropy is written as the difference between the $n \to 1$ replicated polymer partition function with \enquote{point-to-line} and with \enquote{point-to-point} boundary conditions. Here, however, we avoid replicas and consider the distribution of the DP partition function $Z^{(\tau)}$. In general, this is a difficult problem since the configurations entering  $Z^{(\tau)}$ have different degrees of correlation based on the overlap between different paths.
For the tree, this difficulty can be solved using self-similarity as in Ref.~\cite{derrida1988polymers}: we observe that a tree of level $\tau+1$ can be obtained by juxtaposing $K$ independent trees of level $\tau$ and connecting their vertices with a branching point. In terms of the $Z^{(\tau)}$, this leads to the recurrence relation
\begin{equation} \label{eq:rec_Z}
    Z^{(\tau+1)} \stackrel{\text{in law}}{=} B(a)\sum_{\kappa=1}^K Z_\kappa^{(\tau)}\,,
\end{equation}
where the $Z_\kappa^{(\tau)}$'s are $K$ independent realizations of $Z^{(\tau)}$, $a$ is drawn from $P_0(a)$ and the equality is meant in law for probability distributions. 
Eq.\,\eqref{eq:rec_Z} can be turned into a deterministic recursive equation for the Laplace transform $G_\tau(y) := \average{\exp(- e^{-y} Z^{(\tau)})}_0$
\begin{equation} \label{eq:G_tau}
    G_{\tau+1}(y) = \average{ G_\tau\left( y - \ln B(a) \right)^K}_{0} \,.
\end{equation}
The evolution equation \eqref{eq:G_tau} belongs to a wide class of nonlinear reaction-diffusion equations including the famous Kolmogorov-Petrovsky-Piskunov (KPP) equation valid for continuous space and time (see below). Quite generally, the solution $G_{\tau}(y)$ behaves as a ballistically moving stationary wavefront, monotonically interpolating between $G_{\tau}(-\infty) = 0$ and $G_{\tau}(\infty) = 1$. This traveling stationary solution characterizes the distribution of $\ln Z^{(\tau)}$ for large $\tau$ around its typical value. In our case, however, it is more convenient to rewrite $G_{\tau}(y) = 1 - e^{-y} \mathsf{u}_{\tau}(y)$. Because of the normalization condition $\average{Z^{(\tau)}}_0 = 1$, $G_{\tau}(y) = 1 - e^{-y} + O(e^{-2y})$ and $\mathsf{u}_{\tau}(+\infty) =1$, while from $G_\tau(y \to -\infty)\to 0$, we deduce $\mathsf{u}_{\tau}(y \to - \infty) \sim e^{y} \to 0$. In terms of this function, we express~(see EM)
\begin{equation} \label{eq:ZlogZ_u}
      \average{Z^{(\tau)}\ln Z^{(\tau)}}_0 = \int_{-\infty}^\infty dy \ \left(  \mathsf{u}_{0}(y) -  \mathsf{u}_{\tau}(y) \right) \, . 
\end{equation}

\begin{figure}
    \centering
    \includegraphics[width=\columnwidth]{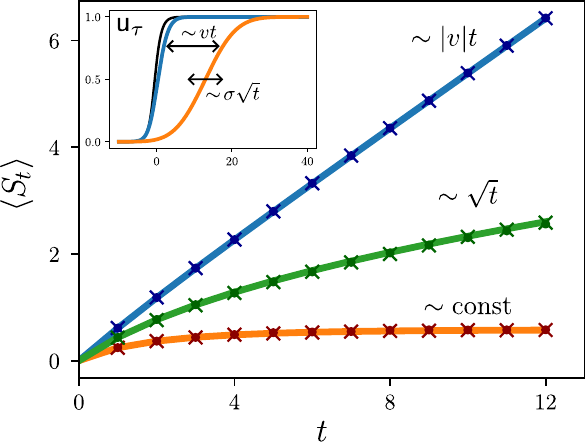}
    \caption{Numerical analysis with $K=2$, $P_0(a) \sim \mathcal{N}(0,1)$  a zero-centered Gaussian, and $P_1(a)\sim \mathcal{N}(\mu, 1)$, so that $D_{\rm}(P_1||P_0) = \operatorname{Var}(P_1||P_0)/2 = \mu^2/2$ \cite{supplnote}. For $v<0$, $\average{S_t}\sim |v| t$ grows linearly in time (blue, $v=-0.51$); for $v>0$, $\average{S_t}$ reaches a constant finite value (orange, $v=0.93$). For $v=0$, $\average{S_t}\propto \sqrt{t}$ (green). 
    Dots and crosses display $\average{S_t}$ from trajectories of the particle $x_\tau$ and of the evolved probabilities \eqref{eq:p_evolve}, respectively~\cite{supplnote}. 
    Full lines show the entropy \eqref{eq:Z_entropy} obtained numerically solving Eqs.~(\ref{eq:G_tau}, \ref{eq:ZlogZ_u}) for $\mathsf{u}_\tau(y)$. 
    Inset: solutions of Eq.~\eqref{eq:KPP_u} are compared to the initial condition (black). For $v>0$ (orange), the drift and broadening is clearly visible. For $v<0$ (blue), $\mathsf{u}_\tau(y)$ converges to a limiting form $e^{y} h_\infty(y)$.
    }
    \label{fig:regimes}
\end{figure}

\textit{Entropy growth rate ---} 
While our model is formulated for $K = e^{\lambda \Delta t} \in \mathbb{N}$, Eq.\,\eqref{eq:G_tau} allows us to take the continuous-time limit $\Delta t \rightarrow 0$ while $\lambda = O(1)$. Consistently, as suggested by the control parameter $v$ defined by Eq.\,\eqref{eq:v_ref}, the accuracy of the measurements has to be scaled 
choosing $P_1 (a) = P_0 (a) + O(\sqrt{\Delta t})$. As shown in the EM, this corresponds to setting $D_{\rm KL} (P_1 \parallel P_0) = \sigma^2 \Delta t/2$, which is the only residual parameter of the finer structure of $P_{0,1}(a)$. Note that for the illustrative Gaussian case this simply corresponds to setting $\mu = \sigma \sqrt{\Delta t}$.
This limit simplifies the discussion from a technical point of view, although it is not strictly necessary as the same phenomenology can be obtained in discrete time (see EM). Setting $G_\tau(y) = 1  - h_t(y)$, in the $\Delta t \to 0$ limit, Eq.\,\eqref{eq:G_tau} assumes the more familiar form of the KPP equation
\begin{equation}
\label{eq:h}
    \partial_t h = \frac{\sigma^2}{2} \partial_y^2 h + \left(\lambda + \frac{\sigma^2}{2}\right) \partial_y h + \lambda F(h)
\end{equation}
where $F(h) = - (1-h)\ln(1-h)$. Beyond this specific form, the results will be universal, given some general properties such as that $F(0) = F(1) = 0$, with $F'(0) = 1$ and $F''(h) < 0$, implying that $h=1$ and $h=0$ are fixed points, respectively stable/unstable. 
At long times there are two cases: for
$\sigma \geq \sqrt{2 \lambda}$ the solution behaves as a traveling wave, $h_t(y) \simeq \bar{h}(y - {\sf y}_t)$ with 
${\sf y}_t \simeq v_{\rm KPP} t - \frac{2 \alpha \sigma}{\sqrt{2 \lambda}} \ln t + o(1)$ and the translation speed $v_{\rm KPP} = -(\frac{\sigma }{\sqrt{2}}-\sqrt{\lambda })^2$; for $\sigma< \sqrt{2 \lambda}$, it converges to
a limit $h_t(y) \to h_\infty(y)$ and $v_{\rm KPP} = 0$~\cite{footnote1}.
However, it is more relevant for us to take the $\Delta t \to0$ limit of Eq.~\eqref{eq:ZlogZ_u} setting $u_t(y) := e^y h_t(y) = \mathsf{u}_{\tau}(y)$
and considering the corresponding partial differential equation
\begin{equation} \label{eq:KPP_u}
    \partial_t u_t (y) = - v \partial_y u_t (y) + \frac{\sigma^2}{2} \partial_y^2 u_t (y) - \lambda e^{y} \tilde{F}(e^{-y} u) \, , 
\end{equation}
where $\tilde F(h) := h - F(h)$ and $v = \sigma^2/2 - \lambda$ comes from the continuum limit of Eq.~\eqref{eq:v_ref}. In the limit of non-informative measurements $\sigma \to 0$, the entropy grows linearly as $\langle S_t \rangle = \lambda t$, hence we first discuss the asymptotic rate of entropy production $s := \lim_{t \to \infty} \langle S_t \rangle/t$. In this perspective, since $u_t(y) e^{-y} \stackrel{y \to -\infty}{\longrightarrow} 1$, for any fixed $y_0$, the contribution for $y < y_0$ to the integral in the r.h.s. of Eq.\,\eqref{eq:ZlogZ_u} is $O(1)$ in time. 
It is thus inessential to the calculation of the rate which is instead controlled by large positive $y$. 
In this regime, $h(y) \ll 1$ and since
$\tilde F(h) = O(h^2)$ at small $h$, we can neglect the non-linear part in Eq.\,\eqref{eq:KPP_u}. As  $u_t$ grows from $0$ to $1$, we can interpret $u_t (y)$ as the cumulative probability distribution of a Wiener process with drift $v$ and diffusion constant $\sigma^2/2$. 
Thus, in this linearized approximation, $u_t(y)$ translates at velocity $v$ while broadening diffusively. For $v \neq 0$, the drift is the dominant factor (see inset in Fig.~\ref{fig:regimes}): for $v>0$, the integral in Eq.\,\eqref{eq:ZlogZ_u} is $\sim v t$; conversely, for $v < 0$ the wavefront exits the domain of integration $y>0$, and $\average{Z^{(\tau)} \ln Z^{(\tau)}}_0 \to O(1)$, (the precise value can be computed in an expansion at small $\sigma^2$, see~\cite{supplnote}). From these considerations, we deduce the exact growth rate 
\begin{equation}
\label{eq:rateres}
s:=\lim_{t\to\infty} \frac{\langle S_t \rangle}{t} = \begin{cases}
  |v| & v\leq 0 \\
  0 & v> 0
\end{cases}     \;.
\end{equation} 
As anticipated, the velocity $v$ tunes a continuous phase transition of the rate of entropy production.  Note that the diffusive front described by Eq.~\eqref{eq:KPP_u} should not be confused with the traveling wave $h_t(y) = \bar{h}(y - {\sf y}_t)$ from \eqref{eq:h}: in $h_t(y)$, the front $u_t(y)$ is visible only as an exponentially suppressed far tail at very large $y$ \cite{supplnote}. In particular, for $\sigma > \sqrt{2 \lambda}$, 
$v_{\rm KPP}<0$ and $v>0$ so that the two fronts move in opposite directions, a manifestation of the fact that $\langle Z^{(\tau)} \ln Z^{(\tau)}\rangle_0$ is controlled by rare instances of $\ln Z^{(\tau)}$. Instead, when $\sigma < \sqrt{2 \lambda}$, $v_{\rm KPP}=0$ and $v<0$: this indicates that the propagation to the left of $u_t(y)=e^{y}h_t(y)$ must eventually stop due to nonlinearity~(see inset in  Fig.~\ref{fig:regimes}), although this has no effect on the growth rate $s$.  Being the Kullback-Leibler divergence positive, in this phase $|v| \leq \lambda$: as $\mathcal{N}_\tau \sim e^{|v| t}$, $|v|$ plays the role of a reduced effective Lyapunov exponent (see Fig.\,\ref{fig:binary_tree}). 
At $\sigma = \sqrt{2\lambda}$, $v=0$, the front \eqref{eq:KPP_u} broadens diffusively, so one expects $\langle S_t \rangle = O(\sqrt{t})$. However, in the critical case, a more careful analysis of nonlinearity is needed as we explain below. 

\begin{figure}
    \centering
    \includegraphics[width=\columnwidth]{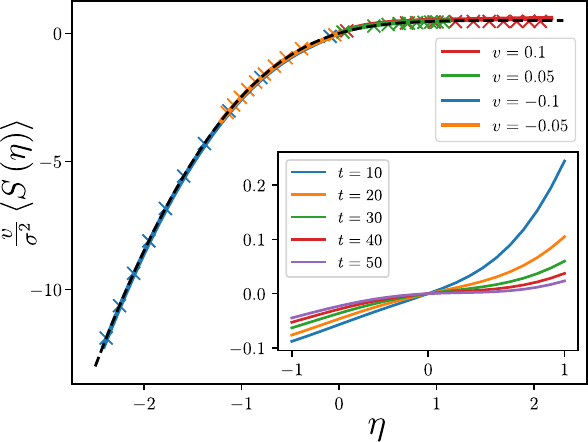}
    \caption{
    Scaling limit for the discrete model (see Caption of Fig.~\ref{fig:regimes}) by solving Eq.~\eqref{eq:G_tau} (crosses) and  for the continuous-time model (full lines) by solving Eq.~\eqref{eq:KPP_u}. In both cases, we compute $v/\sigma^2 \average{S}$ as a function of $\eta = v/\sigma \sqrt{t/2}$, at times up to $t=10^3$, for various values of $v$. Numerical results are compared to the theoretical scaling function $\mathcal{S}(\eta)$ of Eq.~\eqref{eq:scalingfun} (black dashed line).
    Inset: The difference $v/\sigma^2\average{S} - \mathcal{S}$ is shown, as a function of $\eta$ for increasing times $t$.}
    \label{fig:scalingfun}
\end{figure}

\textit{Critical regime ---} 
For small $v<0$, equating $|v| t_v \sim \sqrt{t_v}$, one needs $t \gtrsim t_v = |v|^{-2}$, to distinguish the critical behavior from the linear growth. This suggests to consider the limit $v \rightarrow 0$, $t \rightarrow \infty$ while keeping the rescaled time $T = v^2 t/\sigma^2$ fixed.  
From the diffusive part in Eq.\,\eqref{eq:KPP_u}, one sees that we also have to scale the space variable as $Y = y |v|/\sigma^2$ and consider $U_T(Y) = u_t(y)$.
In the $v\to 0$ limit, for $Y>0$, the nonlinear term becomes negligible as $e^{\sigma^2 Y/|v|} \tilde{F}(e^{-\sigma^2Y/|v|} u) \sim e^{-\sigma^2 Y/|v|} u^2 \rightarrow 0$; on the other hand, for $Y<0$, as $0<h_\tau(y)<1$,
$U_T(Y) < e^{-\sigma^2|Y|/|v|} \stackrel{v\to0}{\longrightarrow} 0$. In other words, $U_T(Y)$
satisfies drifted diffusion for $Y>0$ but with a wall imposing $U_T (Y \leq 0) = 0$. 
Since furthermore $U_T(+\infty)=1$, one can interpret $U_T (Y)$ as the cumulative probability of a drifted Wiener process with a reflecting wall at $Y=0$. Its expression can be computed explicitly (see EM), leading to the %following
asymptotic large time behavior close to criticality
\begin{equation} \label{eq:entropy_scaling}
    \average{S_t} \simeq \frac{\sigma^2}{v} \mathcal{S} \left( v \sqrt{\frac{t}{2 \sigma^2}} \right) 
\end{equation}
with the scaling function
\begin{equation} \label{eq:scalingfun}
    \mathcal{S}(\eta)= \left( \frac{1}{2} + \eta^2 \right) \text{erf} \ \eta - \eta^2 + \frac{\eta}{\sqrt{\pi}} e^{-\eta^2} \ . 
\end{equation}
For $v=0$ this gives $\average{S_t} = \sigma \sqrt{2 t /\pi}$, while in the regimes $\eta \rightarrow \pm \infty$, we recover Eq.~\eqref{eq:rateres} for $v>0$ and $v<0$ respectively. Moreover, we can now compute the critical scaling of $\average{S_t}$, namely $\average{S_t} \sim \sigma^2/(2 v)$ for $v \rightarrow 0^+$ and $\average{S_t} \sim |v| t + \sigma^2/(2 |v|)$ for $v \rightarrow 0^{-}$ consistently with Eq.~\eqref{eq:rateres}.
Note that Eq.~\eqref{eq:entropy_scaling} and the scaling form \eqref{eq:scalingfun} are completely universal in our protocol and also apply to the discrete case where $v$ is given by Eq.~\eqref{eq:v_ref}, while $\sigma^2 = \operatorname{Var}(P_1||P_0)/\Delta t$ with $\operatorname{Var}(P_1||P_0):= \average{\ln^2(P_1(a)/P_0(a))}_1 - \average{\ln(P_1(a)/P_0(a)) }_1^2$. Indeed, a comparison with the numerics performed on the discrete model shows perfect agreement~(see Fig.~\ref{fig:scalingfun}).

\textit{Conclusions ---} In this Letter, we considered the competition between exponential spreading of uncertainty and Bayesian updating of information by repeated measures. We introduced a toy model in terms of the directed random walk of a particle on a tree and obtained a connection to the DP on the Cayley tree. We could then employ the many tools available for this framework, but with significant differences due to the reweighting of polymer configurations due to Bayes' theorem. Our results can be regarded both as a simple and solvable example of a MIPT for a classical particle on the tree and as a fascinating transition in chaos mitigation. 
Interestingly, we find that the critical point for the MIPT discussed here coincides with the well-known freezing transition of the DP, even though the MIPT is dominated by rare events and the critical properties are indeed different. 

Several perspectives open up. From the practical standpoint of chaos mitigation, it would be of interest to analyze more optimized strategies where one tries to minimize the number of measurements to be taken while still pinpointing the state of the system. From the perspective of the DP, it would be interesting
to assess whether the coincidence of the two critical points (MIPT and freezing) observed on the tree is a more general property. Additionally, while no transition is expected for a lattice in dimension $d=1$~\cite{PhysRevLett.129.260603}, the tree provides a good qualitative description for sufficiently high $d$ (presumably for $d > 2$ when the polymer shows a high temperature phase
and self-averaging properties \cite{halpin1995kinetic,comets2017directed}).  

\textit{Acknowledgements ---}
FG acknowledges support from Université Paris-Saclay.
GG and ADL acknowledge support by the ANR JCJC grant ANR-21-CE47-0003 (TamEnt).
PLD acknowledges support from ANR grant ANR-23-CE30-0020-01 EDIPS. GG acknowledges the support of the MSCA Grant 101152898 (DREAMS).

\bibliography{biblio}

\clearpage 

\appendix
\setcounter{equation}{0}
\setcounter{figure}{0}
\renewcommand{\thetable}{S\arabic{table}}
\renewcommand{\theequation}{S\thesection.\arabic{equation}}
\renewcommand{\thefigure}{S\arabic{figure}}
\setcounter{secnumdepth}{2}

\begin{center}

{\Large End Matter \\ 
\vspace{0.22cm}
}
\end{center}

\section{One-point contribution to $\average{S_t}$}
We will now compute the one-point contribution to $\average{S_t}$, i.e., the second term in the r.h.s. of Eq.\,\eqref{eq:Z_entropy}. This can be rewritten in terms of replicas as 
\begin{equation}
    \sum_{j=1}^{K^\tau} \average{z_j^{(\tau)}\ln z_j^{(\tau)}}_0 = \partial_n \sum_{j=1}^{K^\tau} \left. \average{\left(z_j^{(\tau)}\right)^n}_0 \right|_{n =1} \, , 
\end{equation}
while, according to our definition of the $z_j$ 
\begin{equation}
    z_j^{(\tau)} = \prod_{p \in \rm{branch}} \frac{P_1(a_p)}{K P_0(a_p)} \ , 
\end{equation}
where the product runs over the tree branch that connects site $j$ to the origin. As all the $a_p$ are independent one has
\begin{equation}
    \sum_{j=1}^{K^\tau} \average{\left(z_j^{(\tau)}\right)^n}_0 = K^{-(n-1) \tau} \average{ \left(\frac{P_1 (a)}{P_0(a)} \right)^{n-1}}_1^\tau
\end{equation}
where we made use of the fact that $\average{(P_1(a)/P_0(a))^n}_0 = \average{(P_1(a)/P_0(a))^{n-1}}_1$. Finally, by taking the derivative we find
\begin{equation}
    \sum_{j=1}^{K^\tau} \average{z_j^{(\tau)}\ln z_j^{(\tau)}}_0  = \tau \left(D_{\rm KL} (P_1 \parallel  P_0) - \ln K \right) \, . 
\end{equation}

\section{Proof of Eq.\,\eqref{eq:ZlogZ_u}}
We want now to estimate the collective contribution to $\average{S_t}$, i.e. the first term in the r.h.s. of Eq.\,\eqref{eq:Z_entropy}. First, we express it in terms of $G_{\tau} (y)$. To do so we notice that, integrating twice in $Z$ both sides of the identity 
\begin{equation}
    Z^{-1} = \int_0^{+\infty} ds \, e^{-sZ} 
\end{equation}
we get
\begin{equation}
    Z \ln Z = \int^{\infty}_0 \frac{ds}{s^2} \ \left( e^{-s Z} - e^{-s} + (Z-1) e^{-s} s(s+1) \right)  
\end{equation}
Let us now set $Z=Z^{(\tau)}$, $s= e^{-y}$ and take the average of both sides: by taking into account the fact that $\average{Z^{(\tau)}}_0 = 1$ and $G_0(y) = e^{-e^{-y}}$, we get 
\begin{equation}
    \average{Z^{(\tau)}\ln Z^{(\tau)}}_0 = \int_{-\infty}^\infty dy \ e^{y} \left(  G_{\tau}(y) -  G_{0}(y) \right) \, .
\end{equation}
that, expressed in terms of $\mathsf{u}_\tau (y) = e^{-y} \left(1 - G_\tau(y) \right)$ gives Eq.\,\eqref{eq:ZlogZ_u}.

\section{Discrete-time case}
We will show that we can recover the linearized form of Eq.\,\eqref{eq:KPP_u} without the assumption of small $\Delta t$, in the regime $\tau \gg 1$. Indeed, by expressing the discrete recursion relation Eq.\,\eqref{eq:G_tau} in terms of $\mathsf{u}_{\tau} (y) = e^{y} \left(1 - G_\tau(y) \right)$ and expanding to the leading order in $\mathsf{u}_{\tau} (y) e^{-y}$ we get
\begin{equation} \label{eq:linear_u}
    \mathsf{u}_{\tau +1} (y) = \average{\mathsf{u}_{\tau} \left(y + \ln K - \ln \frac{P_1(a)}{P_0(a)} \right)}_1 \, , 
\end{equation}
where now $\mathsf{u}_0 (y) = e^y (1- e^{-e^{-y}})$. The solution can thus be expressed as
\begin{equation}
    \mathsf{u}_{\tau} (y) = \int dy^{\prime} \ \mathcal{U}_\tau (y^{\prime}) \mathsf{u}_{0} (y + y^{\prime}) 
\end{equation}
where $\mathcal{U}_{\tau} (y)$ is the Green's function associated to Eq.\,\eqref{eq:linear_u}. The latter can in turn be expressed in terms of Fourier modes as 
\begin{equation}
    \mathcal{U}_\tau (y) = \int \frac{dq}{2 \pi} e^{i q y + \omega(q) \tau}
\end{equation}
with $ \omega(q) = i q \ln K + \ln \average{e^{-iq \ln P_1(a)/P_0(a)}}_1$. We now want to evaluate the latter expression in the large-time limit. Setting $q \rightarrow q/\sqrt{\tau}$, only the first two orders in $q$ of $\omega(q)$ contribute: as $\omega(q) = - i q v \Delta t -  \text{Var}(P_1 \parallel P_0) q^2/2 + O(q^3)$ we see that, for any $t = \tau \Delta t$
\begin{equation}
    \mathcal{U}_\tau (y) = \frac{e^{-(y-v t)^2/(2 \sigma^2 t)}}{\sqrt{2 \pi \sigma^2 t}} + O(\tau^{-1}) \, ,
\end{equation}
and 
\begin{equation}
    \mathsf{u}_{\tau} (y) = \int \ \frac{dy^{\prime}}{\sqrt{2 \pi \sigma^2 t}} e^{-(y-v t)^2/(2 \sigma^2 t)} \mathsf{u}_{0} (y + y^{\prime}) \,, 
\end{equation}
that provides the general solution of Eq.\,\eqref{eq:KPP_u} in the linear regime. 

\section{Continuum limit}
To properly define the continuum limit, one has to take $P_0(a) - P_1(a) \sim O (\sqrt{\Delta t})$, namely
\begin{equation}
    P_0 (a) = P_1 (a) + \pi_1 (a) \sqrt{\Delta t} + \pi_2 (a) \Delta t + O( \Delta t)^{3/2}
\end{equation}
where the normalization requires $\int \pi_{1,2} (a) = 0$. From this one has, up to terms $O( \Delta t)^{3/2}$
\begin{equation} \label{expansion}
    \ln \frac{P_1(a)}{P_0(a)} = - \frac{\pi_1(a)}{P_1(a)} \sqrt{\Delta t} + \left( \frac{1}{2} \frac{\pi_1^2(a)}{P_1^2(a)} - \frac{\pi_2(a)}{P_1(a)} \right) \Delta t \,.
\end{equation}
Taking the average of \eqref{expansion} w.r.t. $P_1$ gives, at leading order
\begin{equation}
    D_{\rm KL} (P_1 \parallel P_0 ) \,  \simeq \frac{\Delta t}{2} \int da \ \frac{\pi_1^2(a)}{P_1(a)} := \frac{\sigma^2}{2} \Delta t \,, 
\end{equation}
while taking the variance of \eqref{expansion} w.r.t. $P_1$ gives 
\bea 
  &&   \text{Var} (P_1 \parallel P_0) = \text{Var} (P_0 \parallel P_1) \\
  &&  = \average{\ln^2(P_1(a)/P_0(a))}_1 - \average{\ln(P_1(a)/P_0(a)) }_1^2 \nonumber  \\
 &&   =  \Delta t \int da \ \frac{\pi_1^2(a)}{P_1(a)} + o(\Delta t) = 2 D_{\rm KL} (P_1 \parallel P_0 ) + o(\Delta t) \ , \nonumber 
\eea
which is the relation presented in the main text, with $\text{Var} (P_1 \parallel P_0) = \sigma^2 \Delta t$.  

Note that to obtain Eq.~\eqref{eq:h} we also used that, to leading order
\be 
\average{\ln \frac{P_1(a)}{P_0(a)} }_0 \simeq - \Delta t \int da \ \frac{\pi_1^2(a)}{P_1(a)} 
= - \frac{\sigma^2}{2} \Delta t \,.
\ee

\section{Rate of entropy production} 
From Eq.\,\eqref{eq:Z_entropy} and \eqref{eq:ZlogZ_u}, one has the exact relation
\be \label{timederivative} 
\partial_t \langle S_t \rangle = - v - \int_{-\infty}^{+\infty} dy \partial_t u_t(y) \,,
\ee 
Using \eqref{eq:KPP_u} and integrating over $y \in (-\infty,\infty)$ using that 
$u_t(y)$ vanishes at $y=-\infty$ and tends to $1$ at $y=+\infty$, one finds
another exact relation for the entropy production at time $t$
\be 
\partial_t \langle S_t \rangle = \lambda \int_{-\infty}^{+\infty} dy  e^{y} \tilde{F}(e^{-y}  u_t(y))  \,.
\ee  
One can check that $h^2/2 < \tilde F(h) < h^2$. Hence one needs to evaluate 
$A_t = \int_{-\infty}^{+\infty} dy  e^{-y} u_t(y)^2 = \int_{-\infty}^{+\infty} dy  e^{y} h_t(y)^2$. 
This is always a convergent integral since $h_t(y) \to 1$ for $y \to -\infty$ 
and $h_t(y) \sim e^{-y}$ for $y \to +\infty$. For $v>0$ the KPP front solution $h_t(y)=\bar h(y- {\sf y}_t)$ 
has a strictly negative velocity, which implies that on $[y_0, +\infty)$ for any $y_0$ one has that $h_t(y) \to 0$. 
Hence for $v>0$, $A_t \to 0$ and the rate of entropy production vanishes $s=0$. More precisely one finds that for $\sqrt{2 \lambda} < \sigma < 2 \sqrt{2 \lambda}$
the decay of $A_t$ is dominated by the KPP front with $A_t \to e^{{\sf y}_t} \int dz e^z \bar h(z)^2 \sim e^{- |v_{\rm KPP}| t}$, 
while for $ \sigma > 2 \sqrt{2 \lambda}$ it is dominated by the far tail of $h_t(y)$. 
For $v=0$ the KPP front moves more slowly to the left, ${\sf y}_t \simeq - \frac{1}{2} \ln t$.
leading to slower decay of $A_t \sim 1/\sqrt{t}$, consistent with the results in the main text.

\section{Critical scaling}
By setting $y = \sigma^2 Y/|v|$, $t = \sigma^2 T/v^2$, Eq. \eqref{eq:KPP_u} becomes, for $U_T(Y)=u_t(y)$ with $Y,T = O(1)$:
\begin{equation} \label{diff2} 
    \partial_T U_T (Y) = - \xi \partial_Y U_T (Y) + \frac{1}{2} \partial^2_Y U_T (Y) + \dots \, , 
\end{equation}
with $\xi = {\rm sign}(v)$ and $U_T ( + \infty) = 1$. The non linear part $\dots$ reads $\frac{\sigma^2}{v^2} e^{\sigma^2 Y/|v|} \tilde{F}(e^{-\sigma^2 Y/|v|} U_T(Y))$.
As explained in the text, it can be neglected for $Y>0$, while it acts as  a wall imposing $U_T(Y\leq0) = 0$. Thus, Eq.~\eqref{diff2} is the evolution equation for the cumulative probability density of a Wiener process on $Y>0$, with drift velocity $\xi$, and a reflecting boundary wall at $Y=0$. 
Taking as initial condition $U_{T=0}(Y)=\theta(Y)$,  the 
solution of this problem can be written as a Galilean transformation of the $\xi=0$ solution obtained through the reflection principle~\cite{gardiner1985handbook}, namely 
\footnote{It is also $U_T(Y) = {\rm Prob}(\min_{0 \leq \tau \leq T} X(\tau)>0)$ for a Wiener process of drift $- \xi$ with $X(0)=Y$.} 
\begin{equation}
 U_T (Y) = \Phi \left( \frac{Y-\xi T}{\sqrt{2T}} \right)  -  e^{2\xi Y} \left[ 1 - \Phi \left( \frac{Y+ \xi T}{\sqrt{2 T}} \right) \right]\,,
\end{equation}
where $\Phi(x) = \left( 1 + \text{erf}(x) \right)/2$ is the cumulative of the Gaussian distribution with variance $1/2$. 

To compute the entropy, we use \eqref{timederivative}.
Recalling that $t=\sigma^2 T/v^2$
and $y=\sigma^2 Y/|v|$, this leads to
\be
 \frac{|v|}{\sigma^2}   \partial_T \average{S_T} 
 =  - \xi  - \int_0^{+\infty} dY \ \partial_T U_T (Y)  = \frac{1}{2} \partial_Y U_T (Y)|_{Y=0} \nonumber 
\ee 
implying that 
\be 
\frac{v}{\sigma^2}   \partial_T \average{S_T} =  \frac{e^{-\frac{T}{2}}}{\sqrt{2 \pi }
  \xi  \sqrt{T}}-\frac{1}{2} 
   \text{erfc}\left(\frac{\xi 
   \sqrt{T}}{\sqrt{2}}\right)
\ee
which, integrated over time 
gives Eq.\,\eqref{eq:entropy_scaling} with the scaling function \eqref{eq:scalingfun}
with $\eta = \xi \sqrt{T/2}$, which is analytic in $\eta$. 

Although this derivation was carried out in the continuous model, let us notice that it is possible to retrieve the same result in the discrete-time case $\Delta t=O(1)$ as well. Indeed, rescaling $t=\tau \Delta t = T \sigma^2/v^2$ the variable $T=O(1)$ is naturally continuous via the simultaneous limit $\tau\to\infty$, $v\to 0$. Namely, finite increments $\tau\to \tau+1$ correspond to infinitesimal increments $T\to T+v^2/\sigma^2$.

\newpage

\onecolumngrid

\renewcommand{\thetable}{S\arabic{table}}
\renewcommand{\theequation}{S\thesection.\arabic{equation}}
\renewcommand{\thefigure}{S\arabic{figure}}
\setcounter{secnumdepth}{2}

\begin{center}
{\Large SUPPLEMENTAL MATERIAL}
\end{center}
\begin{center}
\vspace{0.8cm}
{\Large Measurement-Induced Phase Transition in State Estimation of Chaotic Systems and the Directed Polymer}
\end{center}

\begin{center}
Federico Gerbino,$^{1}$ Guido Giachetti,$^{2}$ Pierre Le Doussal,$^{2}$ and Andrea De Luca$^{3}$
\end{center}
\begin{center}
$^1${\em Laboratoire de Physique Théorique et Modèles Statistiques, Université Paris-Saclay, CNRS, 91405 Orsay, France}\\
$^2${\em Laboratoire de Physique de l'\'Ecole Normale Sup\'erieure, CNRS, ENS $\&$ PSL University, Sorbonne Universit\'e, Universit\'e Paris Cité, 75005 Paris, France}\\
$^3${\em Laboratoire de Physique Th\'eorique et Mod\'elisation, CY Cergy Paris Universit\'e, \\
\hphantom{$^\dag$}~CNRS, 95302 Cergy-Pontoise, France}\\
\end{center}

\section{Averaging over realizations of the physical particle \label{sec:avemany}}
In the main text we have formulated our model as follows
\begin{itemize}
    \item A single particle undergoes a directed diffusion on the Cayley tree and we denote as $x_\tau \in \{1,\ldots, K^{\tau}\}$ its position at each time;
    \item At each time-step $\tau$ and each site $j = 1, \ldots, K^\tau$ a measurement is performed resulting in a measurement outcome $a_j^{(\tau)}$ distributed according to $P(a_j | x) := P_{\delta_{j,x}
    }(a)$ (to clarify the notation, we use $\avec^{(\tau)}$ to denote the outcomes of measurements at time-step $\tau$, while $\avec^{(\leq\tau)}$ denotes the collection of all measurement outcomes up to time $\tau$);
    \item the observer uses the knowledge of all $a$'s up to time $\tau$ to update with Bayes' theorem their knowledge of the location of the particle which results in the probabilities $p_j^{(\tau)} = P(j^{(\tau)} | \avec^{\leq(\tau)}) $ that $x_\tau = j$.
\end{itemize}
We are then interested in computing averages over the realizations of the trajectory of the particle of functionals  $\langle F(\pvec^{(\tau)}) \rangle_x$ of the probabilities assigned by the observer at a given time $\tau$. We will now show that one can practically disregard the evolution of the physical particle and  use (2) to evolve the probabilities $\pvec$ and the measurement outcomes $\avec$. 
Since we are dealing with the Cayley tree, we note that there is a unique trajectory ending on each leaf labeled by $x_\tau = 1, \ldots, K^{\tau}$. Given the realization $x_\tau$, the probabilities of the $a$'s factorizes
\begin{equation}
\label{eq:afact}
    P(\avec^{(\leq\tau)} | x_\tau) = \prod_{\tau'\leq\tau}\prod_{j=1}^{K^{\tau'}} P(a_j^{(\tau')} | x_{\tau'}) \,.
\end{equation}
Also, from the knowledge of the $a^{(\tau+1)}$
and of the previous set of probabilities $\pvec^{(\tau)}$, the observer can compute the probabilities at time $\tau+1$
from Eq.~(2), which we can compactly rewrite as
\begin{equation}
P(j^{(\tau+1)} | \avec^{(\tau+1)}, \pvec^{(\tau)}) := \omega_j(\pvec^{(\tau)}, \avec^{(\tau + 1)}) \,.
\end{equation}
Iterating this equation, one can express explicitly the probability assigned by the observer given all the measurement outcomes $\avec^{(\leq\tau)}$
\begin{equation}
\label{eq:Pxa}
P(j^{(\tau)} | \avec^{\leq(\tau)}) := 
 \omega_j( \omega(\ldots \omega(\pvec^{(0)}, \avec^{(1)}), \avec^{(2)}) \ldots \avec^{(\tau)}) = \ldots =: \Omega_j(\avec^{(\leq\tau)}) \,,
\end{equation}
which defines implicitly the function $\Omega$ of all the measurement outcomes at all times $\leq \tau$. By Bayes' theorem, one can also write this more explicitly as
\begin{equation}
\label{eq:OmegaP}
    \Omega_j(\avec^{(\leq\tau)}) =  \frac{P(\avec^{(\leq\tau)} | j^{(\tau)})}{\sum_{j'=1}^{K^{\tau}}P(\avec^{(\leq\tau)} | j^{\prime (\tau)})} \,.
\end{equation}

Combining \eqref{eq:afact} and \eqref{eq:Pxa}, we can write the probability distribution for the $p_j$'s assigned by the observer given a realization of the particle
\begin{equation}
\label{eq:Ppx}
    P(\pvec^{(\tau)} | x_\tau) := \int d\avec^{(\leq\tau)} 
    \delta(\pvec^{(\tau)} - \Omega(\avec^{(\leq\tau)}))    P(\avec^{(\leq\tau)} | x_\tau) \,.
\end{equation}
Then, averaging over all trajectories of the physical particle, we get the distribution we are interested in
\begin{equation}
\label{eq:Pp}
    P(\pvec^{(\tau)}) = \frac{1}{K^{\tau}}\sum_{x_\tau} P(\pvec^{(\tau)} | x_\tau) \,.
\end{equation}
Using Eq.~\eqref{eq:Pxa} and \eqref{eq:OmegaP}, we obtain that
\begin{equation}
\label{eq:Ppx1}
    P(\pvec^{(\tau)} | x_\tau) := p^{(\tau)}_{x_\tau} \int d\avec^{(\leq\tau)} 
    \delta(\pvec^{(\tau)} - \Omega(\avec^{(\leq\tau)}))     \sum_{j'}P(\avec^{(\leq\tau)} | j^{\prime (\tau)}) = K^{\tau} p^{(\tau)}_{x_\tau} P(\pvec^{(\tau)})
\end{equation}

We can now show that Eq.~\eqref{eq:Pp} admits a recursive representation. Indeed,
\begin{multline}
    P(\pvec^{(\tau+1)}) = \frac{1}{K^{\tau + 1}}\sum_{x^{(\tau+1)}}
    \int d\avec^{(\leq\tau+1)} 
    \delta(\pvec^{(\tau+1)} - \Omega(\avec^{(\leq\tau+1)}))    P(\avec^{(\leq\tau+1)} | x_{\tau+1}) = \\
    \frac{1}{K^{\tau + 1}}\sum_{x_{\tau+1}}
    \int d\avec^{(\tau+1)} \int d\pvec^{(\tau)}
    \delta(\pvec^{(\tau+1)} - \omega(\pvec^{(\tau)},\avec^{(\tau+1)})) 
    P(\avec^{(\tau+1)} | x_{\tau+1}) 
    \times \\ 
    \int d\avec^{(\leq\tau)}
    \delta(\pvec^{(\tau)} - \Omega(\avec^{(\leq\tau)})   P(\avec^{(\leq\tau)} | x_{\tau+1}) 
\end{multline}
Now, the end point $x_{\tau+1}$ fixes also the trajectory at $x_\tau$ and in the last line we recognise \eqref{eq:Ppx}. Thus from Eq.~\eqref{eq:Ppx1}, we arrive at
\begin{equation}
    P(\pvec^{(\tau+1)}) = 
    \int d\avec^{(\tau+1)} \int d\pvec^{(\tau)}
    \delta(\pvec^{(\tau+1)} - \omega(\pvec^{(\tau)},\avec^{(\tau+1)}))
    P(\pvec^{(\tau)})
    \sum_{x_{\tau+1}}
    P(\avec^{(\tau+1)} | x_{\tau+1}) 
    \frac{p^{(\tau)}_{x_\tau}}{K}
\end{equation}
and using that $p^{(\tau)}_{x_\tau}/K = p^{(\tau+1,-)}_{x^{(\tau + 1)}}$, we precisely obtain that the $p_j$ can be evolved using (2), where the $\avec^{(\tau+1)}$ are extracted ignoring the position of the physical particle, following:
\begin{equation} \label{eq:a_distrib}
    P(\mathbf a^{(\tau+1)} |\mathbf p^{(\tau+1,-)} ) = \sum_{j}P(\mathbf a^{(\tau+1)}|j) \, p_j^{(\tau+1,-)} \,.
\end{equation}

Moreover, from the above representation we have that averages $\average{ F(\pvec^{(\tau)})}$ can be conveniently expressed as 
\begin{equation} \label{eq:ave_new}
    \average{ F(\pvec^{(\tau)})} = \int d\pvec^{(\tau)} F(\pvec^{(\tau)}) P(\pvec^{(\tau)}) = \int d\avec^{(\leq \tau)} F(\Omega(\avec^{(\leq \tau)})) \prod_{\tau'=1}^{\tau} P(\avec^{(\tau')} | \Omega(\avec^{(\leq\tau')})) \,.
\end{equation}

\section{Equivalence of the two averages} 
\label{sec:equiv}

In the main text, we have expressed the probability distribution $P(\avec^{(\tau)}|\pvec^{(\tau,-)})$ of measurement outcomes $\avec^{(\tau)}$ at time-step $\tau$, conditioned on the probabilities $\pvec^{(\tau,-)}$ before measurements (c.f. Eq.~\eqref{eq:a_distrib}). We now show that the same quantity can be written in terms of the unnormalized variables $\mathbf{z}^{(\tau,-)}$.
Starting from Eq.~\eqref{eq:a_distrib} one has
\begin{equation} 
    P(\avec^{(\tau)}|\pvec^{(\tau,-)}) = \sum_{j=1}^{K^\tau} P(\avec^{(\tau)}|j)\,p_j^{(\tau,-)} = \sum_{j=1}^{K^\tau} P(\avec^{(\tau)}|j)\,\frac{z_j^{\tau,-}}{\sum_{j'=1}^{K^\tau} z_{j'}^{\tau,-}} \,.
\end{equation}
Then, exploiting the fact that 
\begin{equation}
    \sum_{j=1}^{K^\tau} z_j^{\tau,-} = \sum_{j=1}^{K^\tau} \frac{z_{\lceil j/K \rceil}^{\tau-1}}{K} = K \sum_{j=1}^{K^{\tau-1}}  \frac{z_j^{\tau-1}}{K} = Z^{(\tau-1)} \,,
\end{equation}
inserting the definition for $P(\avec^{(\tau)}|j)$ (Eq.~(1) of the main text) and using that $P_1(a_j^{(\tau+1)}) / P_0(a_j^{(\tau+1)}) z^{(\tau,-)}_j = z^{(\tau)}_j$, one finds the simpler expression:
\begin{equation} \label{eq:telescopic}
    \sum_{j=1}^{K^\tau} P(\avec^{(\tau)}|j)\,\frac{z_j^{\tau,-}}{\sum_{j'=1}^{K^\tau} z_{j'}^{\tau,-}} = \frac{\prod_{j=1}^{K^\tau} P_0(a_j^{(\tau)})}{Z^{(\tau-1)}} \sum_{j=1}^{K^\tau} \frac{P_1(a_j^{(\tau)})}{P_0(a_j^{(\tau)})} z_j^{\tau,-} =  \frac{Z^{(\tau)}}{Z^{(\tau-1)}} \prod_{j=1}^{K^\tau} P_0(a_j^{(\tau)})  \,,
\end{equation}
where the last term $\prod_{j=1}^{K^\tau} P_0(a_j^{(\tau)})$ is the probability distribution for the unbiased process where all $a_j^{(\tau)}$ are independent with distribution $P_0(a)$.
Let us notice that the ratio $Z^{(\tau)}/Z^{(\tau-1)}$ factorizes at each $\tau$. Therefore, when taking products of distributions $P(\avec^{(\tau')}|\pvec^{(\tau',-)})$ at increasing time-steps $1\leq \tau'\leq\tau$, as in the r.h.s. of \eqref{eq:ave_new}, only $Z^{(0)}=1$ and $Z^{(\tau)}$ do not cancel out, yielding:
\begin{equation} \label{eq:equivv}
    \prod_{\tau'=1}^\tau P(\avec^{(\tau')}|\pvec^{(\tau',-)}) = Z^{(\tau)} \prod_{\tau'=1}^\tau \left( \prod_{j=1}^{K^{\tau'}} P_0(a_j^{(\tau')}) \right) \,.
\end{equation}
The product in the above line corresponds exactly to the probability distribution of the full set of outcomes $\avec^{(\leq \tau)}$. Finally, expressing functionals of probabilities $F[\{\pvec^{(\tau)}\}]$ as functionals of the $\mathbf{z^{(\tau)}}$, from Eqs. (\ref{eq:ave_new}, \ref{eq:equivv}) one precisely obtains the equivalence between the two averages in Eq.~(4) of the main text, namely:
\begin{equation}
\begin{split}
    \average{F[\{\pvec^{(\tau)}\}]} & = \int d\avec^{(\leq \tau)}  \prod_{\tau'=1}^{\tau} P(\avec^{(\tau')} | \Omega(\avec^{(\leq\tau')}))
    F(\Omega(\avec^{(\leq \tau)}))
    = \\
    & = \int d\avec^{(\leq\tau)} \prod_{\tau'=1}^\tau \left( \prod_{j=1}^{K^{\tau'}} P_0(a_j^{(\tau')}) \right) F\left[\left\{\frac{z^{(\tau)}_j}{Z^{(\tau)}}\right\} \right] Z^{(\tau)}
    = \average{F\left[\left\{\frac{z^{(\tau)}_j}{Z^{(\tau)}}\right\}\right]\,Z^{(\tau)}}_0 \,.
\end{split}
\end{equation}

\section{Recall some details on the properties of the KPP equation} \label{sec:recallKPP} 
\subsection{Velocity of the front}

Consider the continuum model (10). The front velocity is obtained as follows [65].
One inserts $h_t(y) \simeq \bar{h}(z=y - {\sf y}_t)$, assuming $\lim_{t \to + \infty} \frac{d}{dt} {\sf y}_t = v_{\rm KPP}$ in (10) leading to
\be 0 = \frac{\sigma^2}{2}  \bar h''(z) + 
(v_{\rm KPP} + \lambda + \frac{\sigma^2}{2} ) \bar h'(z)  + \lambda F(\bar h(z)) 
\ee 
which determines $\bar h(z)$ if $v_{\rm KPP}$ is known. One then focuses on
the forward region $z \gg 1$, inserting $\bar h(z) \sim e^{- \mu z}$ 
to linear order one finds the equation which determines $v_{\rm KPP}$
as a function of $\mu$
\be 
\tilde v_{\rm KPP} = v_{\rm KPP} + \lambda + \frac{\sigma^2}{2} 
=
\frac{\sigma^2}{2} \mu + \frac{\lambda}{\mu}
\ee 
This parabola has a minimum at $\mu=\mu_c= \sqrt{2 \lambda}/\sigma$. Since the initial condition decays as
$h_0(y) \sim e^{-y} $ for $y \to +\infty$ one finds that 
\bea  
&& \sigma < \sqrt{2 \lambda} \quad , \quad \mu=1 \quad , \quad v_{\rm KPP}= 0 \\
&& \sigma > \sqrt{2 \lambda} \quad , \quad \mu=\mu_c \quad , \quad v_{\rm KPP}= - (\sqrt{\lambda}- \frac{\sigma}{\sqrt{2}})^2 \nonumber
\eea  
Here $v_{\rm KPP}= \lim_{t \to +\infty} 
 \frac{1}{t} \ln Z^{(\tau)}$ is the intensive free energy associated
to typical polymer paths (i.e. $n=0$). The first line corresponds to the high temperature phase of the DP and the second to the
low temperature phase where the front velocity is frozen (more precisely it is $\tilde v_{\rm KPP}$ which freezes
to which one must add the drift $- (\lambda + \frac{\sigma^2}{2})$, which corresponds to an additional
energy cost proportional to the polymer length. More precisely one has, following [65]

(i) for $\sigma < \sqrt{2 \lambda}$, ${\sf y}_t \simeq O(1)$ and the
KPP front decays as $\bar h(z) \sim e^{-z}$

(ii) for $\sigma > \sqrt{2 \lambda}$, 
${\sf y}_t \simeq v_{\rm KPP} t -  \frac{\alpha}{\lambda} \tilde v_{\rm KPP}  \ln(\lambda t) + O(1)$ 
where $\tilde v_{\rm KPP} = \sigma \sqrt{2 \lambda}$, with
$\alpha=3/4$, and $\bar h(z) \sim z e^{-\mu_c z}$ for $z \gg 1$
with $\mu_c= \sqrt{2 \lambda}/\sigma$. At the transition $\sigma = \sqrt{2 \lambda}$ the same
holds with 
$\alpha=1/4$ leading to ${\sf y}_t \simeq - \frac{1}{2} \ln t + O(1)$.

\subsection{Discrete model}
Consider now the discrete time model. The linearized form of the recursion Eq.~(8)
with $G_\tau(y)=1-h_\tau(y)$ reads 
 \begin{equation} \label{eq:linear_h2}
    h_{\tau +1} (y) = K \average{h_{\tau} \left(y - B(a)  \right)}_0 \, , 
 \end{equation}
 where $B(a)=- \ln K + \ln \frac{P_1(a)}{P_0(a)}$.
Looking for a front solution $h_\tau(y)=\bar h(y- c \tau)$ with $\bar h(z) \sim e^{- \mu z}$ we find
\be 
c = c(\mu) = \frac{1}{\mu} \ln \left(   \average{K e^{\mu B(a)}  }_0  \right) 
\ee 
In the high temperature phase of the polymer $\mu=1$ and 
\be 
v_{\rm KPP} = \frac{c(1)}{\Delta t} = \frac{1}{\Delta t}  \ln\left( \average{K e^{B(a)}  }_0  \right)  = 0 
\ee 
In the low temperature phase of the polymer the front velocity and the parameter $\mu=\mu_c$ 
are determined by the conditions 
\bea 
&& v_{\rm KPP} = c(\mu_c) = \frac{1}{\mu_c} \ln\left(   \average{K e^{\mu_c B(a)}  }_0  \right)  \\
&& \partial_\mu c(\mu)|_{\mu = \mu_c} = 0 
\eea 
more precisely $\mu_c$ realizes the  minimum of the function $c(\mu)$.
The transition occurs when $\mu_c=1$ and one can check that it corresponds to
\bea 
- \ln K + \average{ \frac{P_1(a)}{P_0(a)}  \ln \frac{P_1(a)}{P_0(a)}  }_0 = v \Delta t = 0 
\eea 
Thus it is a general property that the entropy rate transition occurs at the same location as the
freezing transition of the DP. 

\section{Weak noise expansion}

In the weak noise/high temperature phase, i.e. $v<0$, $\sigma < \sqrt{2 \lambda}$, the solution of the KPP equation converges at large time to a limit, $h_t(y) \to h_\infty(y)$.
One can compute this limit in a systematic weak noise/high temperature expansion. 
Setting $z= e^{-y}$, we look for a stationary solution of (10) (i.e. setting $\partial_t h_t(y)=0$ there) in the form, 
\be \label{series} 
h_\infty(y) = 1 - e^{-z} - Q(z) e^{-z}   \, , \quad Q(z)=\sum_{n \geq 1} \sigma^{2 n} Q_n(z)  \,,
\ee 
where the coefficients, which must obey $Q_n(z)=O(z^2)$ for small $z$, are found as polynomials in $z$. 
Inserting \eqref{series} into (10) one finds
\be 
Q(z) = \frac{\sigma ^2 z^2}{2 \lambda } +\frac{\sigma ^4
   (z-2)^2 z^2}{8 \lambda
   ^2} + \frac{\sigma ^6 (z-2) z^2
   ((z-6) (z-4) z-12)}{48
   \lambda ^3} + O(\sigma^8) \,,
\ee 
leading to 
\be
\lim_{t \to +\infty} ( \langle S_t \rangle + v t ) = \lim_{\tau \to +\infty} \average{Z^{(\tau)}\ln Z^{(\tau)}}_0 
= \int_0^{+\infty} \frac{dz}{z^2} Q(z) e^{-z} = \frac{\sigma
   ^2}{2 \lambda }  + \frac{\sigma ^4}{4
   \lambda ^2} + 
\frac{\sigma ^6}{12 \lambda
   ^3} + O(\sigma^8) \,.
\ee

\begin{figure}
    \centering
    \includegraphics[width=0.45\columnwidth]{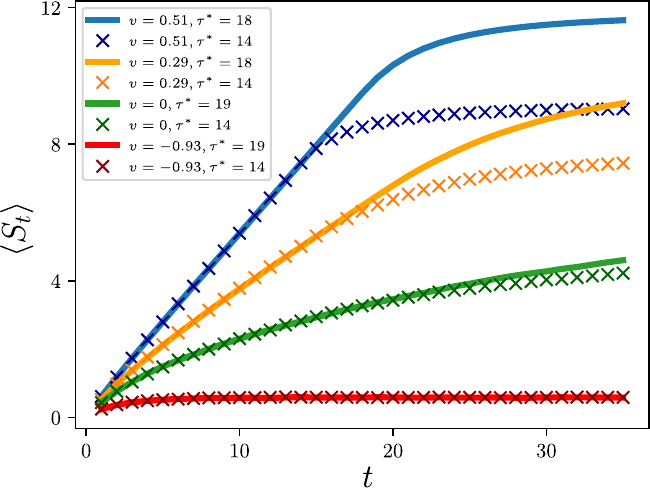}
    \hspace{0.05\columnwidth}
    \includegraphics[width=0.45\columnwidth]{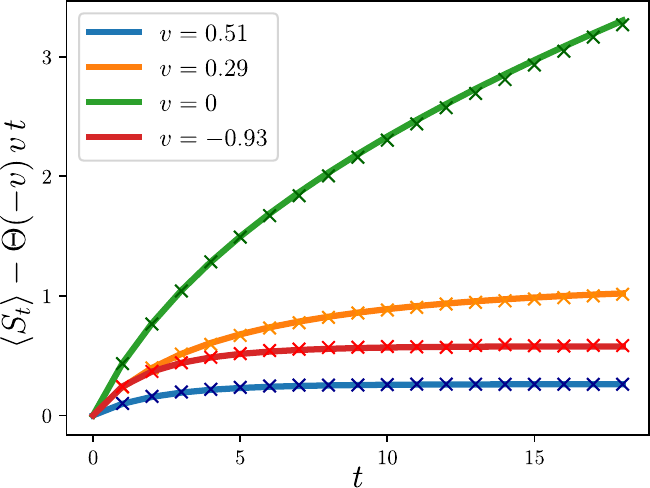}
    \caption{
    \textit{(Left:)} averaged entropy $\average{S_t}$, for various values of the control parameter $v$. Trajectories are obtained evolving the biased probabilities $\pvec$. For small truncation times $\tau^*$ and $\mu<\mu_c$, the entropy $\average{S_t}$ saturates to a value $\sim \tau^*$. Conversely, when $\mu>\mu_c$, the entropy $\average{S_t}$ can be computed exactly at all times, undependently on the value of $\tau^*$. 
    \textit{(Right:)} finite contributions to the entropy $\average{S_t}$. For negative values of $v$, we subtract the leading terms $\sim |v|t$. Markers represent data from simulations of the exact dynamics, while full lines are obtained via  numerical solutions of Eq.~\eqref{eq:u_KPP_2}. The curves show good agreement between the two numerical approaches and the analytical results presented in the main text.  
    }
    \label{fig:simulations}
\end{figure}

\section{Numerical simulations}

In this Section, we provide details about the numerics presented in the main text.

\subsection{Montecarlo dynamics of the particle on the tree}
First, we carry out simulations of the physical single-particle hopping process on a binary Cayley tree $K=2$, choosing $P_0\sim \mathcal{N}(0,1)$ a Gaussian distribution, and $P_1\sim \mathcal{N}(\mu,1)$ a shifted Gaussian averaging to $\mu$. With this choice of the probability distributions our control parameter becomes $v = D_{\rm KL}(P_1 \parallel P_0) - \ln K = \mu^2/2-\ln 2$,  and the critical value of $\mu$ corresponding to $v=0$ is then $\mu_c=\sqrt{2\ln 2}$. 
We follow two equivalent approaches:
\begin{enumerate}
\item We can pick a random trajectory $x_\tau$ uniformly distributed among all those on the tree. Accordingly, we can determine the probability distribution for the measurement outcomes $\mathbf{a}^{(\tau)}$ simply using the conditional probability rule Eq.~(1) of the main text. Since all physical trajectories are statistically equivalent, we can evolve $p_j^{(\tau)}$ considering $x_\tau=0\;\forall \tau$. 
\item Alternatively, as explained in the main text and in Sec.~\ref{sec:avemany}, we can use  Eq.~\eqref{eq:a_distrib}  to generate the measurement outcomes $\mathbf{a}^{(\tau)}$ from the known probabilities $\mathbf{p}^{(\tau)}$ and Eq.~(2) to consequently update the probabilities $\mathbf{p}^{(\tau)} \to \mathbf{p}^{(\tau+1)}$ themselves. 
\end{enumerate}
As a benchmark, we show the agreement between these two simulation protocols in Fig.~2 of the main text.

In both cases, because of the exponential growth $2^\tau$ of the number of leafs in the tree, the simulability of the dynamics is restricted to few iterations. In order to bypass this difficulty, we apply a truncation protocol: after $\tau^*$ exact iterations, we only consider the $2^{\tau^*-1}$ highest probabilities to generate the $2^{\tau^*}$ new probabilities for the subsequent time-step. Enforcing normalization $\sum_{j=0}^{2^{\tau^*}-1} p_j^{(\tau)}=1$, the latter are used to compute the entropy $S_t$ of the trajectory at time $t=\tau \Delta t$.
In the low-measurement regime $v<0$, where all sites are roughly equiprobable, the truncation saturates the entropy growth to a value $\average{S_{\tau>\tau^*}} \sim \ln 2 \tau^*$ proportional to the cutoff $\tau^*$.
Conversely, in the strong-measurement regime $v>0$, the few highest probabilities are enough to compute the relevant contribution to the entropy $S_t$, which we expect to be constant in time: in this case, truncating the full set of $p_j$'s to the highest $2^{\tau^*}$ values captures the exact dynamics of $\average{S_{\tau>\tau^*}}$, provided $\tau^*$ is large enough. 
In Fig.~\ref{fig:simulations} we show the behavior of $\average{S_t}$ obtained by simulating the protocol as described above,
for various values of $\mu$.

\subsection{Numerical solution of Eq.~(8)}
We numerically estimate the behavior of the term $\average{Z\ln Z}_0$ in Eq.~(5) of the main text, solving numerically the recursive equation (8) for its generating function. More explicitly, for $K=2$, we evolve the equation 
\begin{equation} \label{eq:u_KPP_2}
    \mathsf{u}_{\tau+1}(y) = \average{\mathsf{u}_{\tau} \left(y + \ln 2 - \ln \frac{P_1(a)}{P_0(a)} \right)}_1 - \frac{1}{2} \average{e^{-(y + \ln 2 - \ln P_1(a)/P_0(a))} \mathsf{u}^2_{\tau} \left(y + \ln 2 - \ln \frac{P_1(a)}{P_0(a)} \right)}_1 \,,
\end{equation}
which is obtained, in the binary tree case $K=2$, writing $G_\tau(y) = 1-e^{-y}\mathsf{u}_\tau(y)$. Let us note that the above equation reduces to Eq.~(SC.1) when neglecting the quadratic term on the r.h.s.. 
At each time-step $\tau$, we interpolate the function $\mathsf{u}_\tau$ and evaluate it on the shifted positions $y + \ln 2 - \ln P_1(a)/P_0(a)$, where samples $a$'s are drawn from a Gaussian probability distribution of mean $\mu$ and variance 1. 

\begin{figure}
    \centering
    \includegraphics[width=0.45\columnwidth]{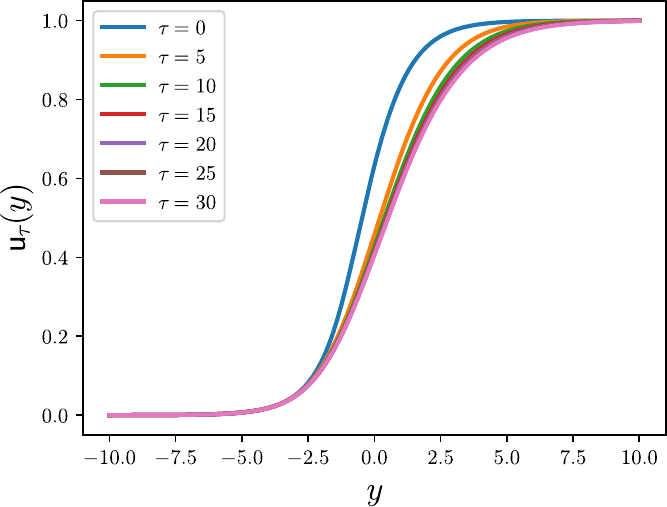}
    \hspace{0.05\columnwidth}
    \includegraphics[width=0.45\columnwidth]{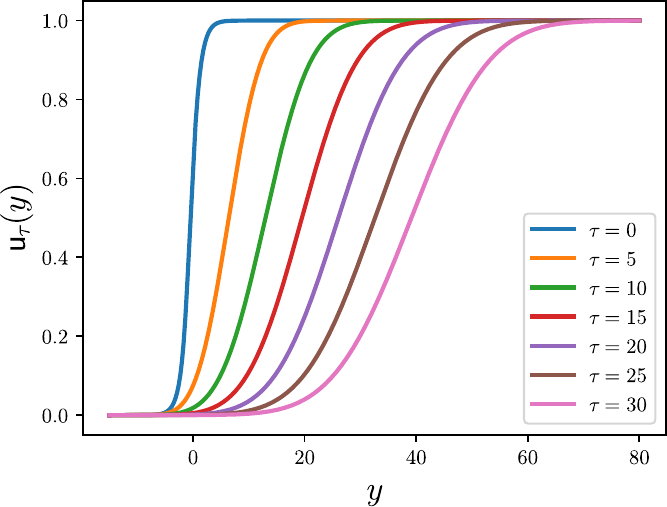}
    \caption{Comparison between the evolved $\mathsf{u}_\tau(y)$ for $v=-0.3$ (left) and $v=1.3$ (right). In the former case, $\mathsf{u}_\tau(y)$ freezes around a limiting shape, whereas in the latter case $\mathsf{u}_\tau(y)$ is a wavefront moving rightwards with velocity $v$. 
    }
    \label{fig:u_wave}
\end{figure}

For $\mu<\mu_c$, corresponding to $v<0$, the function $\mathsf{u}_\tau(y)$ attains a limiting shape, shown in Fig.~\ref{fig:u_wave}, left panel, corresponding to a finite contribution to the entropy through the integral (9). Conversely, for $\mu>\mu_c$ and $v>0$, $\mathsf{u}_\tau(y)$ is a wavefront shifting rightwards with velocity $v$, as displayed in Fig.~\ref{fig:u_wave}, right panel. Its contribution to the entropy is thus of order $\sim vt$ and cancels out with the one-point terms, yielding $\average{S_t}\sim O(1)$. The latter are displayed on the right panel of Fig.~\ref{fig:simulations}, showing complete agreement between the numerical solutions of Eq.~\eqref{eq:u_KPP_2} and simulations of the dynamics.

\end{document}